\documentclass[particles,article,accept,moreauthors,pdftex]{Definitions/mdpi} 
\usepackage{isotope}
\usepackage{physics}
\def\openone{\leavevmode\hbox{\small1\kern-3.8pt\normalsize1}}
\def\beq{\begin{equation}}
\def\eeq{\end{equation}}
\def\be{\begin{eqnarray}}
\def\ee{\end{eqnarray}}

\def\veck{{\bf k}}
\def\vecq{{\bf q}}

\def\magq{|{\bf q}|}

\firstpage{1} 
\pubvolume{1}
\issuenum{1}
\articlenumber{0}
\pubyear{2023}
\copyrightyear{2023}
\externaleditor{Academic Editor: Armen Sedrakian} 
\datereceived{ } 
\daterevised{ } 
\dateaccepted{ } 
\datepublished{ } 
\hreflink{https://doi.org/} 

\Title{Testing the Paradigm of Nuclear Many-Body Theory}

\TitleCitation{Testing the Paradigm of Nuclear Many-Body Theory}

\Author{{Omar Benhar \orcidA{}} 
}

\AuthorCitation{Benhar, O.}

\address[1]{%
INFN and Departnent of Physics, Sapienza University, 00185 Rome, Italy; omar.benhar@roma1.infn.it}

\abstract{Nuclear many-body theory is based on the tenet that nuclear systems can be accurately described as 
collections of point-like particles. This picture, while providing a remarkably accurate explanation of a wealth 
of measured properties of atomic nuclei, is bound to break down in the high-density regime, 
in which degrees of freedom other than protons and neutrons are expected to come into play.
Valuable information on the validity of the description of dense nuclear matter in terms of 
nucleons, needed to firmly establish its limit of applicability, can be obtained from 
electron--nucleus scattering data at large momentum transfer and 
low energy transfer. 
The emergence of $y$-scaling in this kinematic region, unambiguously showing that the beam particles couple to 
high-momentum nucleons belonging to strongly correlated pairs, 
indicates that at densities as large as five times nuclear density\textemdash typical of the neutron star interior\textemdash  nuclear matter largely behaves as a collection of nucleons.
}


\begin{document}
\section{Introduction}
\label{intro}

The available empirical information on nuclear properties demonstrates that in spite of the finite size 
and complex internal structure of protons and neutrons,  ato\-mic nuclei largely behave as collections of point-like constituents
obeying the laws of nonrelativistic quantum mechanics. Investigation of the limits of applicability of this picture, providing the conceptual
basis of nuclear many-body theory, 
is of paramount importance to the development of a unified framework 
for the description of all nuclear systems, from the deuteron to heavy nuclei and neutron stars~\cite{bob_rmp}.

In recent years,  astrophysical observation\textemdash notably the detection of gravitational radiation emitted by a coalescing binary 
neutron star system reported by the LIGO/Virgo Collaboration~\cite{LIGOScientific:2018hze,LIGOScientific:2017vwq,LIGOScientific:2020aai} and the 
mass-radius measurements performed by the NICER satellite~\cite{Cromartie:2019kug,Fonseca:2021wxt,Riley:2019yda,Miller:2019cac,Riley:2021pdl,Miller:2021qha}\textemdash 
have provided unprecedented information that allows constraining the theoretical models of neutron star  matter.
Complementary access to the properties of dense matter can be gained from the analysis of the large database of high-quality electron--nucleus scattering data, 
spanning a broad kinematical region and nuclear targets ranging from deuteron and helium to nuclei as heavy as gold; for a review, see, e.g., Ref.~\cite{benhar_RMP}. 

Experiments in which a beam of weakly interacting particles is scattered off a composite system have long been recognised as
a powerful tool to reveal the internal structure of the  target. These studies exploit the observation\textemdash based on general quantum mechanical considerations\textemdash that a simple two-body reaction mechanism, involving only 
the beam particle and one of the target constituents, becomes dominant at large momentum transfer. As a consequence, the target response in this regime exhibits a remarkable scaling behaviour in a variable simply related to the 
momentum of the struck constituent. The occurrence of scaling is largely independent of the target internal dynamics and has been observed in a variety of different processes, such as neutron scattering off quantum liquids~\cite{he}, electron--nucleus scattering~\cite{nuclei}, and electron--proton scattering ~\cite{bj}. The connection between scaling in many-body systems and Bjorken scaling in deep inelastic scattering has been analysed in Ref.~\cite{BENHAR2000131}.  

In electron--nucleus scattering, the emergence of scaling in the variable $y$ indicates that the beam particles interact 
with point-like target constituents having mass equal to that of nucleons and carrying momenta up to  
500 MeV and above~\cite{yscaling_Arrington}. High-momentum components in the nuclear wave function originate from strong dynamical correlations,  
which give rise to virtual scattering processes leading to excitation of the participating nucleons to states above the Fermi sea.  
Ample experimental evidence of correlations in nuclei has been provided by measurements the cross sections of nucleon knock-out reactions~\cite{benhar_NPN}. The results of these experiments show that correlation effects account for $\sim$20\% of the normalisation of the ground-state wave functions~\cite{benhar_NPN}. 

The appearance of high-momentum nucleons belonging to strongly correlated pairs brings about fluctuations of the matter density $\varrho$, 
the amplitude of which can be estimated using the simple relation linking density and Fermi momentum.
According to this relation, in isospin-symmetric matter, a nucleon Fermi momentum of 500 MeV corresponds to a density 
exceeding the central density of atomic nuclei, $\varrho_0 \sim 0.16 \ {\rm fm}^{-3}$, by a factor of about~seven.

Complementary information on fluctuations of nuclear density induced by short-range nucleon--nucleon correlations has been recently obtained from 
the studies of two-nucleon knock-out processes \isotope[12]{C}$(e,e^\prime p p)$ and \isotope[12]{C}$(e,e^\prime p n)$ carried out at the Thomas Jefferson National Accelerator Facoloty (Jefferson Lab).
The results discussed in Ref.~\cite{Subedi_2008}, suggesting that the density of correlated pairs inferred from the Jefferson Lab data
can be as high as $\sim$5$\varrho_0$, turn out to be largely consistent with those obtained from $y$-scaling analyses.

This article is aimed at providing a short but self-contained introduction to electron--nucleus scattering processes 
involving high-momentum nucleons, whose investigation has provided 
valuable information on the applicability of the paradigm of nuclear many-body theory to describing matter in the density regime
relevant to neutron stars. This information, which should be seen as complementary to that obtained from astrophysical observations, 
is critical to firmly establish the occurrence of a phase transition involving the appearance of quark matter in the neutron 
star core~\cite{Baym:2017whm}.  

The body of the paper is organised as follows. 
In Section~\ref{emergence}, the mechanism driving the emergence of scaling in the dynamic response of many-body systems is analysed 
using the simple case of neutron scattering off liquid helium as a pedagogical example. 
The generalisation to the case of electron--nucleus scattering, involving nontrivial issues associated both with the nature of the 
electromagnetic interaction and with the complexity of nuclear dynamics, is discussed in Section~\ref{IA}. Selected results of scaling analyses
of nuclear data are illustrated in Section~\ref{data}.
Finally, the significance of $y$-scaling for identification of the relevant degrees of freedom in dense nuclear matter and the 
implications for the description of neutron star properties are outlined in Section~\ref{summary}.



\section{Emergence of \boldmath{$y$}-Scaling in the Response of Many-Body Systems}
\label{emergence}

Let us consider, as a pedagogical example,  scattering off a nonrelativistic bound system consisting of N
point-like scalar particles of mass M. Under the assumption that the projectile--target interaction is weak, the differential cross section
of the process in which a beam particle of momentum ${\bf k}$ and
energy $E$ is scattered into the solid angle $d\Omega$ with energy
$E^\prime = E - \omega$ and momentum ${\bf k}^\prime = {\bf k} - {\bf q}$ can be written in Born approximation as
\beq
\frac{d\sigma}{d\Omega d\omega} = \frac{\sigma}{4 \pi}\
\frac {|{\bf k}^\prime|}{|{\bf k}|}
S({\bf q},\omega)\ .
\label{xsec:1}
\eeq

In the above equation, 
$\sigma$ is the total cross section  of the elementary process involving the projectile particle and a target 
constituent. The information on the structure and dynamics of the target is contained in the {\it response function} $S({\bf q},\omega)$\textemdash also referred to as dynamic structure function\textemdash  defined as
\begin{align}
\label{response:12}
S({\bf q},\omega)  & =  \frac{1}{N} \sum_n | \langle n |\rho_{\bf q}| 0 \rangle |^2
\delta(\omega+E_0+E_n) \\ 
\nonumber
&= \frac{1}{{\rm N}}\ \int  \frac{dt}{2\pi}\ {\rm e}^{i\omega t}
\langle 0 | \rho^\dagger_{\vecq}(t)\rho_{\vecq}(0) | 0 \rangle \ .
\end{align}

Here $| 0 \rangle$ and $| n \rangle$ are the target
ground and final states satisfying the Schr\"odinger equations 
$H | 0 \rangle=E_0| 0 \rangle$ and $H | n \rangle=E_n| n \rangle$, with 
$H$ being the target Hamiltonian, and the sum is extended to the complete set of final states.
The time evolution of the operator describing the transition to a state of momentum ${\bf q}$ is dictated by $H$ according to\begin{align}
 \rho_{\bf q}(t) = {\rm e}^{iHt}\rho_{\bf q}{\rm e}^{-iHt} \ \ \ \  , \ \ \ \ 
\rho_{\bf q} = \sum_{\bf k} a^\dagger_{{\bf k}+{\bf q}}a_{\bf k} \ ,
\end{align}
where $a^\dagger_{\bf k}$ and $a_{\bf k}$ are creation and
annihilation operators of the constituent particles.
The matrix elements appearing in Equation~(\ref{response:12}) can be rewritten in coordinate space using 
\begin{align}
\langle R | \rho_{\bf q} | R^\prime \rangle = \delta( R - R^\prime ) \sum_{i=1}^{\rm N} {\rm e}^{i {\bf q}\cdot{\bf r}_i} 
\end{align}
where $R\equiv\{{\bf r}_1,\ldots,{\bf r}_{\rm N}\}$ specifies the target configuration, and the $3N$-dimensional $\delta$-function 
is defined as $\delta( R - R^\prime )  = \prod_{i=1,N} \delta({\bf r}_i - {\bf r}_i^\prime)$. The resulting expression of the response is
\beq
S({\bf q},\omega)= \frac{1}{{\rm N}}\ \sum_n \left| \int dR\  \langle n | R \rangle
\Big( \sum_{i=1}^{\rm N}
 {\rm e}^{i {\bf q}\cdot{\bf r}_i} \Big) \  \langle R | 0 \rangle \right|^2
\delta(\omega+E_0-E_n)\ ,
\label{resp:coord}
\eeq
$\langle R | 0 \rangle$ and $\langle R | n \rangle$ being the wave
functions of the target initial and final state, respectively.

The above expression greatly simplifies the kinematical region in which the impulse approximation (IA) is expected to be applicable. In the IA regime, 
the space resolution of the beam particles, $\lambda =  2 \pi/|{\bf q}|$, satisfies the relation $\lambda \ll d$, with $d$ being the average separation 
distance between target constituents. As a consequence, the scattering process reduces to the incoherent sum of elementary processes involving individual constituents, with the remaining (N-1) particles acting as spectators. Under the further assumption that all interactions between the struck constituent and the spectators
can be disregarded, one can finally write the response function in the form~\cite{impulse} 
\beq
S({\bf q},\omega) = \int \frac{d^3k}{(2\pi)^3}\ n(\veck)\
\delta \Big( \omega + \frac{ {\bf k}^2 }{ 2M } - \frac{ | {\bf k}+ {\bf q} |^2 }{ 2M } \Big)\ ,
\label{s:nk}
\eeq
where the distribution $n(\veck)$ describes the probability of finding a constituent of momentum ${\bf k}$ in the target ground state. It should be noted that, 
because the momentum distribution is an {\it intrinsic} property of the target, the response function in the IA regime only depends on momentum and
energy transfer through the argument of the energy-conserving $\delta$-function, 
implying that 
\beq
\omega - k_\parallel  \frac{ |{\bf q}| }{M} - \frac{\magq^2}{ 2M } = 0 , 
\label{energy:cons}
\eeq
with $k_\parallel$ denoting the component of ${\bf k}$ parallel to the momentum transfer.



\newpage

The onset of scaling---that is, the observation that, up to a kinematical factor, 
$S({\bf q},\omega)$ becomes a function of a single variable---simply 
reflects the
fact that in the IA regime, in which energy conservation is expressed by
Equation~(\ref{energy:cons}), ${\bf q}$ and $\omega$ are {\it not} 
independent variables. One can then define a {\it scaling variable}
$y=y(\magq,\omega)$ and the associated \mbox{{\it scaling function}} 
\begin{align}
\label{def:Fy}
 F( \magq,y ) = K(\magq,\omega) S({\bf q},\omega) \ , \
\end{align}
such that 
\begin{align}
\lim_{\magq \to \infty} F(\magq,y)  = F(y)  \ .
\end{align}

The definitions of both the scaling variable and the scaling function clearly emerge, and Equation~\eqref{s:nk} is rewritten in the form
\begin{align}
S({\bf q},\omega) = \frac{M}{q}  \ 2 \pi \int_{|k_{min}|}^\infty  k dk n(k) \ ,
\end{align}
with $k=|{\bf k}|$, and
\begin{align}
\label{kmin}
k_{min} = \frac{M}{q} \Big( \omega - \frac{{\bf q}^2}{2M} \Big) \ .
\end{align}

The resulting expressions are
\begin{align}
\label{def:fy}
y = k_{min} \ \ \ , \ \ \ F( |{\bf q}|,y ) = \frac{M}{q} S({\bf q},\omega)  \ .
\end{align}

From Equations~\eqref{energy:cons},~\eqref{kmin}, and ~\eqref{def:fy}, it follows that  $y=0$ corresponds to $\omega = \omega_{QE} = {\bf q}^2/2M$, that is, to elastic scattering on a constituent at rest in free space. Positive and negative $y$, on the other hand, correspond to positive or negative $k_\parallel$, implying in turn 
$\omega >  \omega_{QE}$ or $\omega <  \omega_{QE}$, respectively.

\begin{figure}[H]
\includegraphics[scale=0.75]{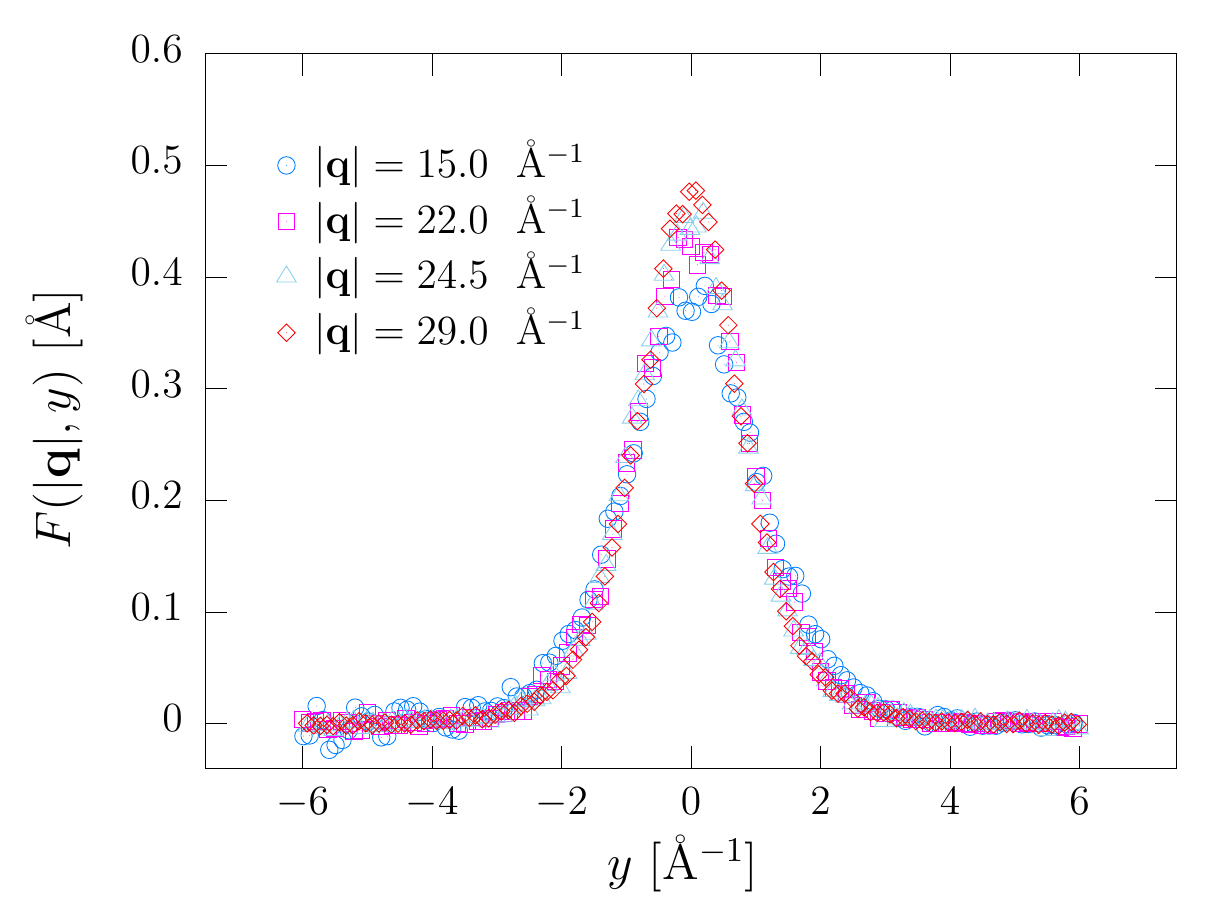}
\caption{Scaling functions $F(y)$, defined as in Equation (\protect\ref{def:fy}), obtained from measurements of neutron 
scattering off superfluid $^4$He at T $=$1.6 K \cite{scal:he}.}
\label{helium}
\end{figure}

Figure~\ref{helium} shows the $q$- and $y$-dependence of the function $F(\magq,y)$ of Equation~\eqref{def:fy} obtained from measurements of neutron scattering off superfluid $^4$He at temperature \mbox{T $=$1.6 K~\cite{scal:he}.} The expected scaling behaviour is clearly observed, for {\it both} positive and negative $y$, at $\magq > 15$ \AA$^{-1}$, while 
appreciable deviations from the IA limit appear at the lowest $\magq$.

It should be pointed out that, while the primary goal of the scaling analysis is showing that scattering off individual 
particles of mass $M$ is the dominant reaction mechanism, a quantitative understanding of scaling violations\textemdash  arising 
mainly from final state interactions (FSI) between the struck atom and the 
spectators\textemdash is also of great importance, because they carry valuable dynamical information. After removal of FSI 
corrections, the relation linking the response in the scaling regime to the distribution $n({\bf k})$ has been extensively exploited 
to obtain the momentum distributions of normal and superfluid \isotope[4][]{He} from neutron scattering data~\cite{PhysRevB.50.6726}.


\section{\boldmath{$y$}-Scaling in Electron--Nucleus Scattering}
\label{IA}

The unpolarised electron--nucleus scattering cross section
is usually written in the form
\beq
\label{A:xsec}
\frac{d^2 \sigma_A}{d\Omega d\omega} = \frac{\alpha^2}{Q^4}\
\frac{E^\prime}{E}\ L_{\mu \nu} W_A^{\mu \nu}\ ,
\eeq
where $\alpha = 1/137$ is the fine structure constant, $E$ and $E^\prime$ denote the initial and final electron energy,
respectively, and $Q^2= - q^2 = {\bf q}^2 - \omega^2$, with ${\bf q}$ and $\omega$ being 
the momentum and energy transfer. 
The tensor $L_{\mu \nu}$ is 
fully specified by the measured lepton kinematic variables, while the target response is described by the tensor
\begin{align}
\nonumber
W^{\mu \nu}_A(q) & = \sum_n \langle 0 | J^\dagger_\mu | n   \rangle \langle n | J_\nu | 0 \rangle
  \delta^{(4)}(P_0+q-P_n) \\
& = \int \frac{d^4x}{(2\pi)^4}\ {\rm e}^{iqx}\
\langle 0 | {J^\mu_A}^\dagger(x) {J^\nu_A}(0) | 0 \rangle\ ,
\label{Wmunu:12}
\end{align}
where $P_0 \equiv (M_A, 0)$\textemdash $M_A$ being the mass of the target nucleus\textemdash and $P_n$ are the four-momenta of the initial and final hadronic states, while $J^\mu_A$ denotes the nuclear electromagnetic current. A comparison between the above definition and Equation~\eqref{response:12} clearly shows the analogy between $W_{\mu\nu}$ and the dynamic structure function $S({\bf q},\omega)$ discussed in the Section~\ref{emergence}.

The starting point for the identification of the scaling variable of electron--nucleus scattering and the corresponding scaling function
is the observation that in the IA regime, the incoming electron interacts with an individual nucleon of momentum ${\bf k}$, while the 
spectator system recoils with energy $E_R = \sqrt{ (M_A - m + E)^2 + {\bf k}^2 }$. In these conditions, the target tensor reduces to a sum of 
contributions arising from protons and neutrons, weighted with the corresponding energy and momentum distributions. The resulting expression reads~\cite{benhar_RMP}
\begin{align}
\label{tens:A}
W_A^{\mu \nu}({\bf q},\omega)  = \int d^3k \ dE \frac{m}{E_{|{\bf k}|}}   \big[ Z w_p^{\mu\nu}(k,{\widetilde q})P_p({\bf k},E) + 
(A-Z) w_n^{\mu\nu} (k,{\widetilde q}) P_n({\bf k},E) \big] \ , 
\end{align}
where $Z$ and $A$ denote the target charge and mass number, $m$ is the nucleon mass, and $k \equiv(E_{|{\bf k}|},{\bf k})$, with $E_{|{\bf k}|} = \sqrt{{\bf k}^2 + m^2}$. 
The spectral functions $P_N({\bf k},E)$, with $N = p, n$, are trivially related to the proton and neutron Green's function, and describe
the probability of removing a nucleon of momentum ${\bf k}$  from the nuclear target, leaving the residual system with energy $E$.
The tensors describing the elementary electromagnetic interactions, involving a {\it moving bound} nucleon, are defined as 
\begin{align}
\label{tens:N}
w_N^{\mu\nu}(k,q) = W_1^N \Big( -g^{\mu\nu} + \frac { {\widetilde q}^\mu  {\widetilde q}^\nu } {\widetilde q^2} \Big)
 + W_2^N  \Big( {k}^\mu - {\widetilde q}^\mu\frac{ ({k} \cdot {\widetilde q}) }{\widetilde q^2} \Big)  
                  \Big( {k}^\nu  - {\widetilde q}^\nu\frac{ ({k} \cdot {\widetilde q}) }{\widetilde q^2} \Big) \ , 
\end{align}
with the metric tensor defined as $g \equiv {\rm diag}(1,-1,-1,-1)$. The replacement of the physical four-momentum transfer with ${\widetilde q} \equiv ({\widetilde \omega},{\bf q})$ 
accounts for the fact that a fraction of the energy transfer goes into the energy of the spectator 
system. This feature becomes manifest in the $|{\bf k}|/m \to 0$ limit, in which the energy transferred to the struck nucleon reduces to 
${\widetilde \omega} = \omega -E$. 
A detailed derivation of Equations~\eqref{tens:A}~and~\eqref{tens:N}  can be found in Ref.~\cite{benhar_RMP}.

The nucleon structure functions  $W_1^N$ and $W_2^N$, extracted from electron--proton and electron--deuteron scattering data, 
depend on ${\widetilde q}^2$ and the squared invariant mass of the hadronic final state produced at the electron-nucleon vertex,  
$W^2 = (k + {\widetilde q})^2$. In the elastic channel, corresponding to $W^2 = m^2$, they can be written in the form
\begin{align}
\label{def:w1}
W_1^N & = -  \frac{ {\widetilde q}^2 }{2m^2} G^2_{M_N}({\widetilde q}^2) \delta \Big( {\widetilde \omega} + \frac{ {\widetilde q}^2 }{ 2m }  \Big)  \ , \\\label{def:w2}
W_2^N & = \Big( 1 - \frac{ {\widetilde q}^2 }{4m^2} \Big)^{-1} 
\Big[ G^2_{E_N}({\widetilde q}^2)  - \frac{ {\widetilde q}^2 }{4m^2} G^2_{M_N}({\widetilde q}^2)  \Big]
\delta \Big( {\widetilde \omega} + \frac{ {\widetilde q}^2 }{ 2m }  \Big) \ , 
\end{align}
where $G_{E_N}$ and $G_{M_N}$ denote the electric and magnetic form factors of the nucleons. 

In the case of elastic scattering, the $\delta$-function appearing in the structure functions\textemdash and, as a consequence, in the nuclear cross section of  Equation~\eqref{A:xsec}\textemdash 
entails a relation between momentum and energy transfer that can be written in the form
\begin{align}
\omega + M_A - \sqrt{ (M_A - m + E)^2 + {\bf k}^2 } - \sqrt{ m^2 + |{\bf k}+{\bf q}|^2 } = 0 \ .
\label{energy_cons:nuclei}
\end{align}

The above equation, analogous to Equation~\eqref{energy:cons}, defines the condition for the appearance of scaling in the variable $y=y(|{\bf q}|,\omega)$ defined through
 the equation~\cite{nuclei}
\begin{align}
\omega + M_A = \sqrt{(y + |{\bf q}|)^2 + m^2 } + \sqrt{ (M_A - m + E_{\rm thr})^2 + y^2}   \ , 
\label{def:y}
\end{align}
where $E_{\rm thr}$ is the threshold for nucleon emission from the target nucleus. It can be easily shown that $y$ is the 
minimum longitudinal momentum carried by a nucleon bound with energy $E_{\rm thr}$; see, e.g., Ref.~\cite{benhar_RMP}.
Note that the variable $y$ is trivially related to the Nachtmann variable $\xi$, which in turn reduces to the Bjorken scaling 
variable of deep inelastic scattering in the $Q^2 \to \infty$ limit~\cite{BENHAR2000131}.

It should be noted that, compared to the case of neutron scattering off liquid helium, the description of electron--nucleus 
scattering involves significant differences. Owing to the vector nature of the electromagnetic coupling, the intrinsic response 
of the target nucleus does not appear as a multiplicative factor in the expression of the electron scattering cross section
 but, rather, as the tensor $W^{\mu\nu}_A$ contracted with the electron tensor $L_{\mu\nu}$; see Equation~\eqref{A:xsec}. 

A second important difference arises from distinct implementations of the assumptions underlying the IA. As pointed out by the authors of Ref.~\cite{impulse}, neglecting altogether the interactions between the struck particle and the 
spectators\textemdash which leads to the appearance of the momentum distribution $n({\bf k})$ in Equation~\eqref{s:nk}\textemdash is fully justified 
in neutron scattering off quantum liquids when the momentum transfer is larger than $\sim$10 \AA, because the energies
involved in these interactions turn out to be negligible with respect to the width of the response function. On the other hand, in electron--nucleus scattering
with momentum transfer around and above $\sim$1 GeV, the excitation energy of the spectator system, described by the nucleon spectral function, can be as high as $\sim$80 MeV, and its effects must be properly taken \mbox{into account.}

Within the IA, a factorised expression of the nuclear cross section\textemdash suitable for the identification of the scaling function\textemdash  can be obtained considering 
that the momentum and energy dependence of the electron--nucleon cross sections is much weaker than that of the spectral functions, 
which turn out to be rapidly decreasing functions of $|{\bf k}|$ and exhibit a pronounced peak at $E=E_{\rm thr}$~\cite{benhar_RMP}. The resulting expression of the scaling function 
turns out to be
\begin{align}
\label{def:lim}
F(y) = \lim_{Q^2 \to \infty} F(y,Q^2) \ , 
\end{align}
where
\begin{align} 
\label{def:fynucl}
F(y,Q^2) = \Big( \frac {d\omega}{d k_\parallel} \Big)_{|{\bf k}|=k_{{\rm min}} }  \frac{ d^2\sigma_A }{   [ Z d^2\sigma_p + (A-Z)d^2\sigma_n ]} , 
\end{align}
with $\sigma_A$ being the {\it measured} electron--nucleus scattering cross section. Here $k_{{\rm min}}$ is the lowest value of $k_\parallel = {\bf k}\cdot{\bf q}/|{\bf q}|$ allowed by kinematics, while $\sigma_p$ and $\sigma_n$ are the {\it elastic} electron--proton and electron--neutron cross 
sections\textemdash stripped of the energy-conserving $\delta$-functions\textemdash evaluated at $|{\bf k}| =   k_{{\rm min}}$ and $E=E_{\rm thr}$.

\section{Observation of \boldmath{$y$}-Scaling in Electron Scattering Data}
\label{data}

The first convincing evidence of $y$-scaling in nuclei has been obtained from the cross sections of the 
process $e+\isotope[3][]{He} \to e^\prime + X$ measured at the Stanford Linear Accelerator Center (SLAC) in the 1970s~\cite{Day:1979bx}.
The SLAC experiment collected data in a broad kinematic region, corresponding to beam energy $3 \lesssim E \lesssim 15$ GeV and 
fixed electron scattering angle $\theta_e = 8$ deg.
Systematic studies, carried out using a variety of targets ranging from  \isotope[4][]{He} to \isotope[197][]{Au}, have been carried out at SLAC~\cite{PhysRevLett.59.427} and
the Thomas Jefferson National Accelerator Facility (Jlab)~\cite{yscaling_Arrington,E12-14-012:inclusive}.

\begin{figure}[H]
\includegraphics[scale=0.80]{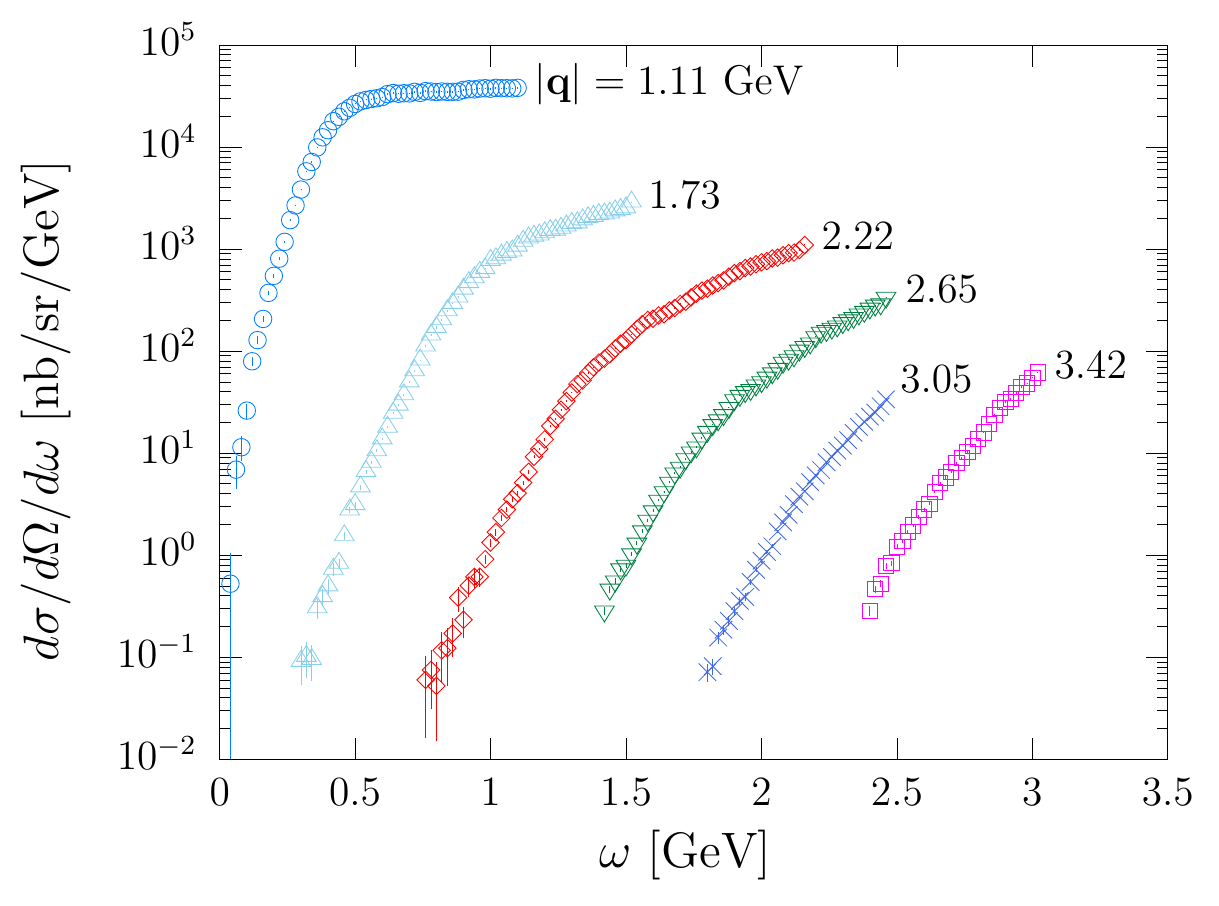}
\caption{Double differential cross section of the process $e + \isotope[197][]{Au} \to e^\prime + X$, 
measured at beam energy $E=4.033$ GeV and electron scattering angles $\theta_e =$ 15, 23, 30, 37, 45, and 55 deg~\cite{yscaling_Arrington}.
The data sets are labelled by the value of the momentum transfer at $\omega = \omega_{QE} = Q^2/2m$.}
\label{Au_dsigma}
\end{figure}

\begin{figure}[H]
\includegraphics[scale=0.80]{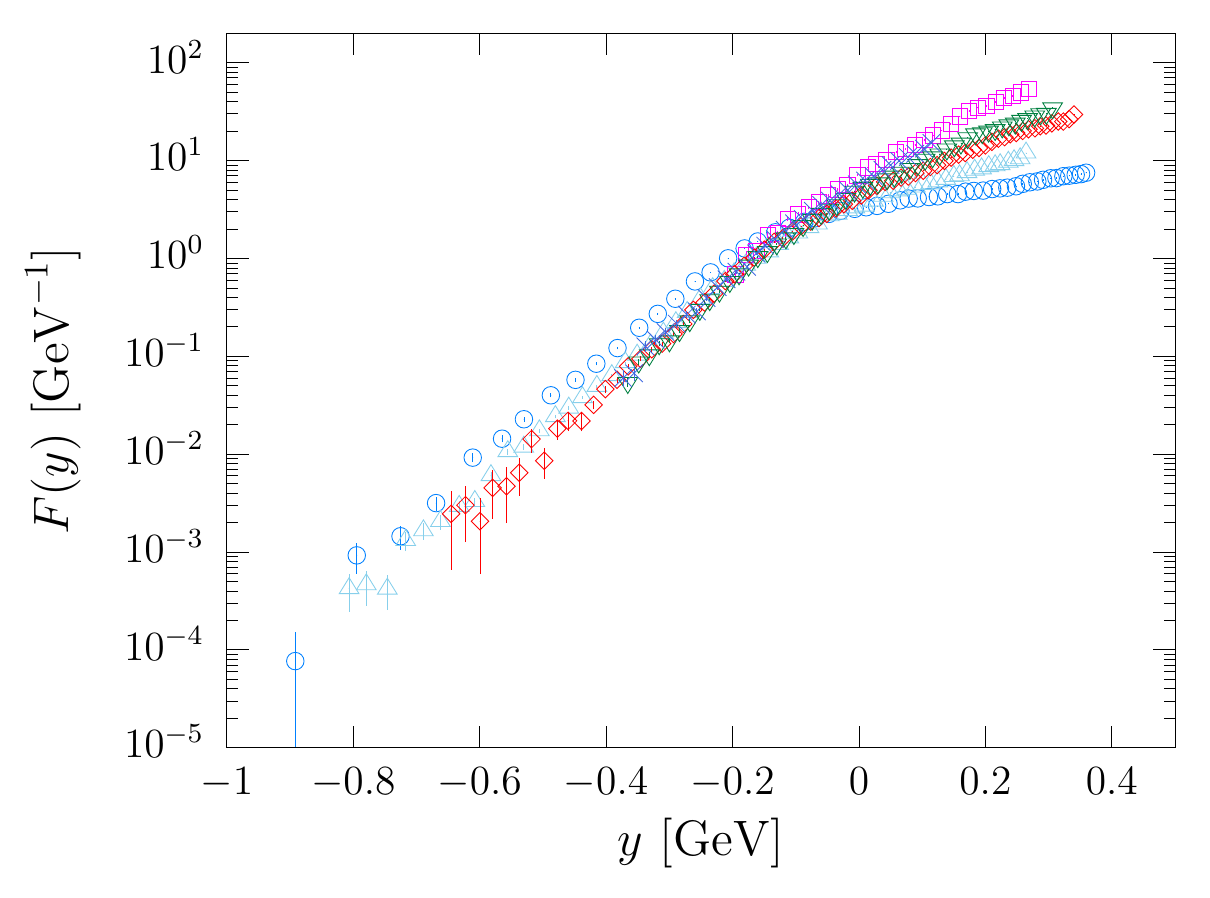}
\caption{Scaling functions $F(y)$, defined as in Equation (\protect\ref{def:fynucl}), obtained from the cross sections displayed in Figure~\ref{Au_dsigma}.}
\label{Au_Fy:y}
\end{figure}

The cross sections of the process $e + \isotope[197][]{Au} \to e^\prime + X$ at incident electron
energy $E=4.033$ GeV and  scattering angles $\theta_e =$ 15, 23, 30, 37, 45, and 55 deg, reported in Ref.~\cite{},  are displayed in Figure~\ref{Au_dsigma}, while Figure~\ref{Au_Fy:y} shows the corresponding scaling functions, defined by Equations~\eqref{def:lim} and \eqref{def:fynucl}.
It is apparent that the cross sections, covering a large kinematic range, extend over many orders of magnitude. On the other hand, the same data shown in terms 
of the scaling functions $F(y,Q^2)$ collapse to a single universal line at {\it negative} $y$.

For any given beam energy and scattering angle, negative $y$ corresponds to energy transfer $\omega~<~\omega_{QE}$, with
$\omega_{QE} = Q^2/2m$ being the energy transfer corresponding to elastic scattering on a free nucleon at rest. As a consequence, in this region 
the incoming electron can only scatter elastically on nucleons having longitudinal momentum $k_\parallel < 0$, and the the condition for the onset
of scaling is fulfilled. On the contrary, scaling is severely violated in the region of positive $y$, where resonance production and deep inelastic scattering\textemdash corresponding to squared invariant mass of the hadronic final state $W^2~>~m^2$\textemdash are the dominant reaction mechanisms. 
 

\begin{figure}[H]
\includegraphics[scale=0.75]{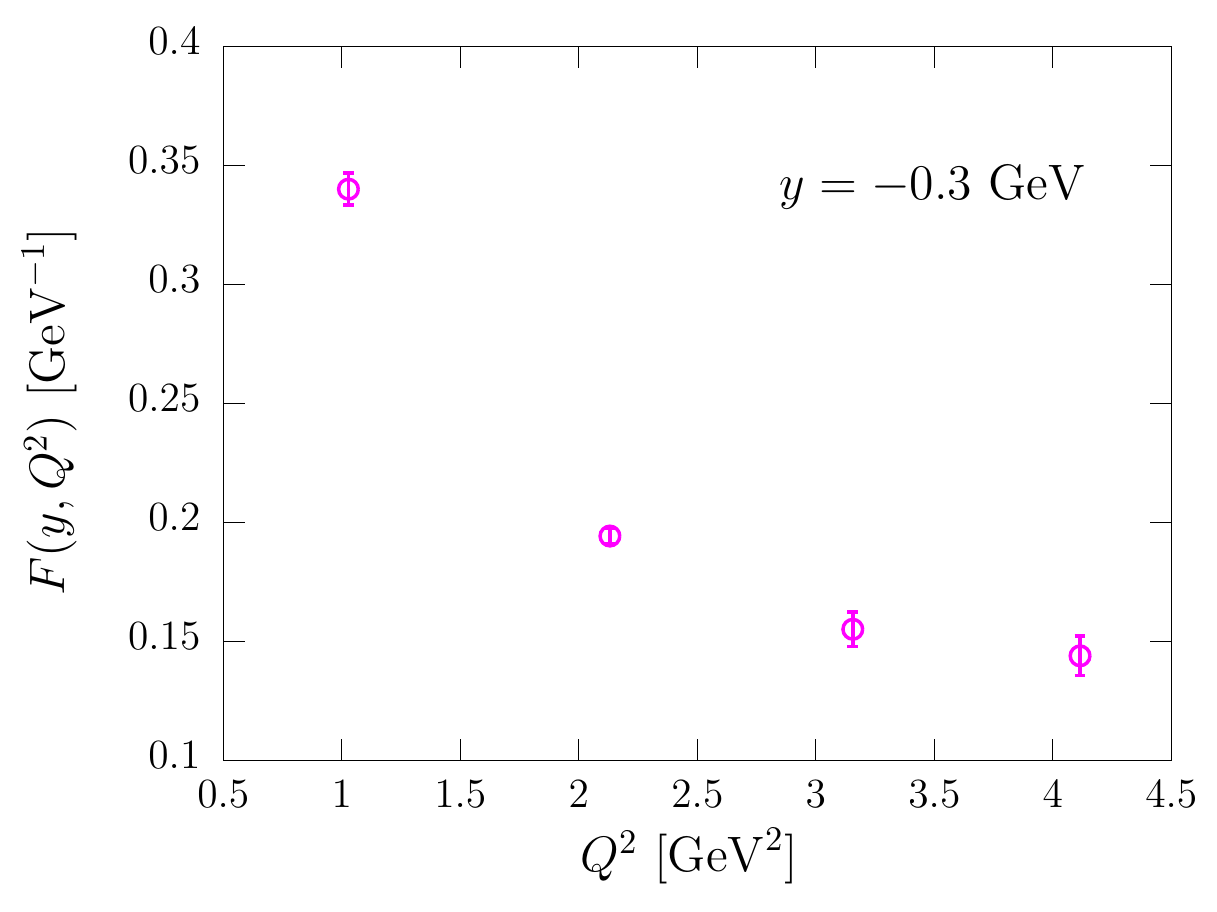}
\caption{$Q^2$-dependence of the scaling function $F(y,Q^2)$ at $y=-0.3$ GeV, obtained from the
cross sections of the process $e + \isotope[197][]{Au} \to e^\prime + X$ reported in Ref.~\cite{yscaling_Arrington}.}
\label{Au_Fy:Q2}
\end{figure}

The small scaling violations at $y<0$ are largely due to FSI and are expected to become vanishingly small in the $Q^2 \to \infty$ limit; see Equation~\eqref{def:lim}.
The $Q^2$-dependence  of the function $F(y,Q^2)$ at $y=-0.3$ GeV is illustrated in Figure~\ref{Au_Fy:Q2}, showing that in the presence of 
FSI, the scaling limit is approached from above and achieved at $Q^2 \gtrsim 4 \ {\rm GeV}^2$~\cite{PhysRevLett.83.3130}.

As already pointed out, in electron--nucleus scattering, the scaling function is defined in terms of spectral functions, not momentum distributions. 
In the case of deuteron, however, the energy dependence of the spectral function reduces to a $\delta$-function, and the 
relation between scaling function and momentum distribution takes the simple  form~\cite{2h_claudio}
\begin{align}
n(k) = -\frac{1}{2\pi} \frac{1}{y} \left. \frac{dF(y)}{dy} \right|_{|y| = k} \ ,
\end{align}
with $k = |{\bf k}|$. The above equation has been employed by the authors of Ref.~\cite{2h_claudio} to obtain the nucleon 
momentum distribution in deuteron from the data reported in Refs.~\cite{PhysRevLett.38.259,PhysRevLett.49.1139}, 
corrected to remove small FSI effects. Besides being valuable in their own right, the results of this study have allowed testing 
the validity of the assumptions underlying the scaling analysis as well as its accuracy.

\begin{figure}[H]
\includegraphics[scale=0.85]{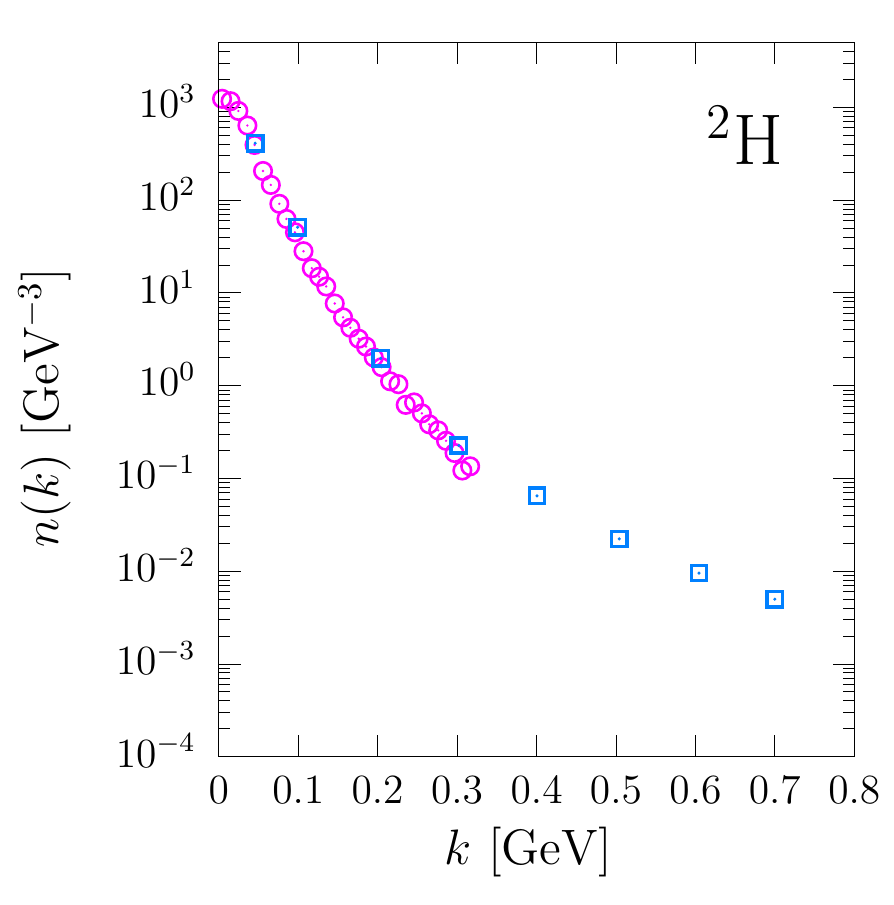}
\caption{Nucleon momentum distribution in deuterons. 
	The squares  represent the results of the $y$-scaling analysis of  Ref.~\cite{2h_claudio}, 
based on the cross sections of the process $e + \isotope[2][]{H} \to e^\prime + X$ reported in Refs.~\cite{PhysRevLett.38.259,PhysRevLett.49.1139}. For comparison, 
the open circles show the results of the analysis of Ref.~\cite{2h_arenhoevel}, based on the \isotope[2][]{H}$(e,e^\prime p)$ data reported in Ref.~\cite{2h_eep}.}
\label{2h:nk}
\end{figure}

In Figure~\ref{2h:nk}, the deuteron momentum distribution of Ref.~\cite{2h_claudio} is compared to the one obtained from the cross section
of the process $\isotope[2][]{H}(e,e^\prime p)n$, in which the scattered electron and the outgoing proton are detected in 
coincidence~\cite{2h_eep,2h_arenhoevel}.
The agreement between the two data sets in the momentum range in which they overlap is striking and indicates that the 
$y$-scaling analysis provides a consistent framework for the determination of $n(k)$ up to momenta as high as 0.7 GeV, well above the region 
accessible by $(e,e^\prime p)$ experiments.  

\section{Summary and Conclusions}
\label{summary}

The $y$-scaling analysis of electron--nucleus cross sections provides ample model-independent evidence that in the kinematic
region corresponding to momentum transfer above $\sim1$~GeV and low electron energy loss, the beam particles interact elastically 
with the protons and neutrons bound in the target nucleus. 

It should be emphasised that the onset of scaling depends not only on the mass of the nucleon\textemdash through the definition of the scaling 
variable $y$ of Equation~\eqref{def:y}\textemdash but also on its electromagnetic properties, described by the structure functions $W_1^N$ and $W_2^N$  
determining the electron--proton and electron--neutron cross sections; see Equation~\eqref{def:fynucl}. 
The results of the study of Ref.~\cite{ingo_PLB}, based on the data reported in Ref.~\cite{Day:1979bx}, show that, in fact,  the 
observation of $y$-scaling sets a 3\% to 6\% limit on the increase of the nucleon size in the nuclear environment.  

The persistence of scaling down to $y=$ $-0.5$   GeV and beyond demonstrates that the electrons interact
with high-momentum nucleons, belonging to strongly correlated pairs, 
at local densities largely exceeding the equilibrium density of isospin-symmetric matter.  This conclusion  is supported by the studies of two-nucleon emission processes carried out at Jefferson Lab. Based on these data, the authors of Ref.~\cite{Subedi_2008} argued that the occurrence of short-range  nucleon--nucleon correlations is associated with strong fluctuations of the nuclear density, which can reach values as high as $\sim$5$\rho_0$. This estimate turns out to be consistent with the observation of scaling at $y \sim~-$0.6 GeV. 

Understanding the density regime in which nucleons are the relevant degrees of freedom is essential to firmly establishing the occurrence of transitions to more exotic forms of matter\textemdash involving baryons other than protons and neutrons as well as deconfined quarks\textemdash which may become energetically favoured at the high densities of the neutron star core. This is a critical issue in view of the results of recent measurements, revealing a small difference between the radii of neutron stars
of mass 1.4 and 2.071~$M_\odot$~\cite{Riley:2019yda,Miller:2021qha}. These data, implying that the equation of state of neutron star matter is still 
rather stiff at $\varrho > 3\varrho_0$, appear to rule out
the occurrence of a strong first-order phase transition in the density range $3 \varrho_0 \lesssim \varrho_B \lesssim 4 \varrho_0$.

As a final remark, it should be pointed out that the applicability of the paradigm underlying nuclear many-body theory at neutron star densities 
paves the way to a novel approach in which astrophysical data can be exploited to test and constrain the existing microscopic models of nuclear dynamics in dense matter~\cite{bayes1,bayes2,rmf0,rmf1,rmf2}.

\funding{This reseacrh was funded by INFN under grant TEONGRAV.}

\dataavailability{This article does not report any new or unpublished data.} 

\conflictsofinterest{The author declares no conflict of interest.}


\begin{adjustwidth}{-\extralength}{0cm}

\reftitle{References}

\PublishersNote{}
\end{adjustwidth}
\end{document}